\documentclass[draft]{agujournal2018}
\usepackage{apacite}
\usepackage{url} 
\usepackage{amsmath}
\usepackage{amssymb}
\usepackage{float}
\usepackage{xcolor}
\usepackage{booktabs}
\usepackage{multirow}
\usepackage{rotating}
\usepackage{graphicx}

\draftfalse

\journalname{Earth and Space Science}

\begin{document}

%
%


\title{Estimating Unknown Cycles in Geophysical Data}

%
%




\authors{Xueheng Shi\affil{1}, Colin Gallagher\affil{1}}

\affiliation{1}{Department of Mathematical Sciences, Clemson University}

\affiliation{1}{O-110 Martin Hall, Box 340975, Clemson, S.C. 29634}




\correspondingauthor{Colin Gallagher}{cgallag@g.clemson.edu}




\begin{keypoints}
\item Geophysical time series tend to have cycles, which can have known or unknown period;
\item Nonparametric statistical methods can estimate unknown period from data in astronomical, geophysical, and climate data.
\end{keypoints}

%
%



\begin{abstract}
Examples of cyclic (periodic) behavior in geophysical data abound.   In many cases the primary period is known, such as in daily measurements of rain, temperature, and sea level.   However, many time series of measurements contain cycles of unknown or varying length.   We consider the problem of estimating the unknown period in a time series.  We review the basic methods, {compare their performance through a simulation study using observed sea level data, apply them to an astronomical} data set, and discuss generalizations of the methods.  
\end{abstract}

%
%

%


\section{Introduction}
Researchers often assume the period of the seasonal component of a time series is known, for example, the annual or diurnal cycle of measurements of surface temperature, precipitation, tides, etc...  However, in some cases, it may be impossible to predetermine the periodicity in a time series. Furthermore, the length of cycles may be changing:  sunspot cycles \citep{kane}, the El Nino-Southern Oscillation \citep{bunge},  or longer term climate cycles \citep{cycle}. {Fourier transform and spectrum analysis have been developed to determine the periodicity for uniformly sampled data.} However, the problem becomes far more complex when the time series is formed by unequally spaced observations, which requires additional research efforts. {Unequally spaced or irregularly sampled time series data exist in many fields. As a motivating example we can consider the study of variable stars,  which has inspired creation of period finding algorithms for unevenly spaced data \citep{reimann1994}.  Here the data are taken at non-uniform times due to the star orbit, weather conditions, and equipment maintenance.  Other examples of unevenly spaced observation arising in geophysical data include paleoclimate proxy records and climate field reconstruction \citep{Mann}, but more commonly unequally spaced data occurs when observations are missing, as in sparse measurements of climatic variables in the early days of instrumentation.}  

{This paper provides a review of some of the statistical methods that have been developed to estimate the {\em unknown} period in a time series. The problem is considered in the general case of unequally spaced observation times-the statistical problem becomes easier when data is at regular intervals.  We review and briefly describe competing statistical methods for determining the unknown period in Section \ref{review}. In Section \ref{compare} we compare the accuracy of the competing methods using sea level data with a known primary period.  In particular, we study the sensitivity of the estimation methods to the amount of data missing (spacing of time points) and to the degree of noise, by randomly removing observations and by adding varying degrees of random noise to the data.  We will see that the Lomb-Scargle periodogram and cubic spline smoothing method are the most robust to missing data (unequal spacing) and to increased noise, in that they are most able to accurately estimate the period regardless of the spacing or noise level.    In section \ref{application} we demonstrate the methods using data on the light magnitude of a variable star.  We end the paper with some discussion of future directions for statistical methods. } 



\section{Literature Review of Nonparametric Period Finding Methods}\label{lit_rev}
 There has been a rich history in spectrum analysis which includes frequency estimations in signal processing. Note the period of a cycle is the reciprocal of the frequency, so for monthly effects with period of 12, the frequency would be 1/12. Conventional spectral analysis techniques like the periodogram requires the input signal to be uniformly sampled which is hardly satisfied in practice. \citet{deeming} began the work on estimation approaches for the unevenly spaced observational data of variable stars. Then dozens of methods have been developed, both parametric and nonparametric. This paper only discusses the nonparametric approaches for period estimation.  

\emph{Periodograms} are generally based on discrete Fourier transform and power spectrum. The classical periodogram was developed more than a century ago by \citet{schuster}. \citet{deeming} applied the discrete Fourier transform to  unequally spaced data in astronomy. \citet{walker} and \citet{hannan} explored the asymptotic properties of the periodogram estimator. Later \citet{lomb}-\citet{scargle} periodogram becomes a well-known and widely used algorithm for periodicity detection in unevenly-spaced time series.   


A second class of approaches search the period by evaluating dispersion either in the sum of lengths between phase-sorted data or sum of dispersion in phased bins compared against trial periods. The String-length method was attributed to \citet{lk}, \citet{renson} and \citet{dwo}. \citet{clarke} presented a generalization ``Rope-length Method'' for multivariate time series data. The Phase Dispersion Minimization (PDM) was due to  \citet{stel}.

The third class of approaches estimate the period by fitting local regression or smoothing splines. Periodic smoothing splines were discussed in the context of spectral estimation by \citet{coda}. The application of cubic spline in the period estimation was introduced by \citet{ake}. \citet{frie} invented a variable-span local linear smoother so called "SuperSmoother", and \citet{reimann1994} first adopted it in his dissertation.

\section{Model and Methodology} \label{review}


Let $g$ be a periodic function with the true period $p_0$. We have $N$ pairs of observations $(t_n, Y_n)$, $1 \le n \le N$, in which $t_n$ represents the time when the observation was made. The observations are ordered by the value of $t_n$, i.e., $0<t_1 \le t_2 \le \cdots \le t_N$. We write the model as
\begin{equation}
Y_n = g(t_n) + \epsilon_n, \qquad 1 \le n \le N
\end{equation}
where $\epsilon_n$'s are {independent identically distributed (i.i.d.)} errors with $E(\epsilon_n|t_n)=0$ and $Var(\epsilon_n|t_n) < \infty $. The typical goals are to estimate $p_0$ and $g(\cdot)$.

A simple idea is to construct a nonparametric estimator $\hat{g}(\cdot|p)$ of $g$ under the assumption that the period of $g$ is $p$. See \citet{hall}. We then map $\hat{g}$ to $\mathbb{R}$ by periodicity and define the Sum Squared Error (SSE) as
\begin{equation}
\text{SSE}(p) = \sum_{n=1}^N \left[ Y_n - \hat{g}(t_n | p) \right]^2.
\end{equation}

We choose the estimator $\hat{p}$ which minimizes $\text{SSE}(p)$. For an appropriate estimator $\tilde{g}(\cdot |p)$ under the assumption of period $p$, we take $\hat{g}=\tilde{g}(\cdot | \hat{p})$ to be the estimator of $g$. This paper will address searching period rather than estimating $g(\cdot)$ in the next few sections.

%
%

\subsection{Fourier Transform based Methods}
\subsubsection {Discrete Fourier Transform, Power Spectrum and Classical Periodogram}
Consider a continuous function $g(t)$ which are uniformly sampled at the discrete time $0<t_1 \le t_2 \le \cdots \le t_N$, the Fourier transform $\mathcal{F}$ for the discrete sampling is
\begin{equation}
\mathcal{F}(g) = \frac{1}{N} \sum_{n=1}^N g(t_n)e^{-i2\pi f t_n},
\end{equation}
where $f$ is the frequency. We use the canonical notation of Fourier Transform in term of the frequency $f$ to keep the consistency with spectrum analysis. Note $f=\frac{1}{p}$.


The squared amplitude, known as the \textbf{power spectrum}, is defined as
\begin{equation}
\mathcal{P}_g = |\mathcal{F}(g)|^2.
\end{equation}


With the \emph{power spectrum} defined, \citet{schuster} first proposed the \textbf{classical periodogram}
\begin{align}
P_S(f) &= \frac{1}{N} \left| \sum_{n=1}^N Y_n e^{-2 \pi i f t_n} \right|^2 \\
&= \frac{1}{N} \left[ \left( \sum_{n=1}^N Y_n \sin (2 \pi f t_n) \right)^2 +  \left( \sum_{n=1}^N Y_n \cos (2 \pi f t_n) \right)^2   \right].
\end{align}

The estimate of the frequency $\hat{f}$ on the interval is the one that maximizes the periodogram $P_S(f)$. 

{However, the classical periodogram has several drawbacks. With unevenly spaced data, the Fourier power spectrum does not have well-defined statistical properties \citep{lomb}, this is because the discrete Fourier transform relies on some strong assumptions: evenly spaced observations of infinite duration, Gaussian white noise, and stationary behavior \citep{vanderplas}. As a consequence, the classical periodogram does not work well on unevenly spaced data, see Table \ref{tab:lcb1} in Section \ref{application}. } 

\subsubsection{Lomb-Scargle Periodogram}

\citet{lomb} and \citet{scargle} proposed a Fourier-like power spectrum estimator to characterize the periodicity in the unevenly spaced data. Lomb-Scargle periodogram can be also seen as fitting the least square of sine waves to the unevenly spaced data,
\begin{align}
    \min_{f}\; \sum_{n=1}^N \left [Y_n - A_f \sin(2\pi f (t_n - \phi_f)) \right]^2, 
\end{align}
where the amplitude $A_f$ and phase $\phi_f$ depend on the trial frequency $f$. With some calculus, the least square solution, so called \textbf{Lomb-Scargle periodogram} is established as
\begin{equation} \label{eq:9}
P_{LS}(f)=\frac{1}{2} \left[
\frac{\sum_{n=1}^N Y_n\cos(2\pi f (t_n - \tau) )  }{\sum_{n=1}^N \cos^2(2\pi f (t_n - \tau) )} + \frac{\sum_{n=1}^N Y_n\sin(2\pi f (t_n - \tau) )  }{\sum_{n=1}^N \sin^2(2\pi f (t_n - \tau) )}
\right],
\end{equation}
where $\tau = \frac{1}{4\pi f} \tan^{-1}  \displaystyle\left[ \frac{\sum_{n=1}^N \sin(4\pi f t_n) }{\sum_{n=1}^N \cos(4 \pi f t_n)} \right ] $.

{\citet{vanderplas} has an in depth discussion on the connection between classic periodogram and Lomb-Scargle periodogram. If the data is evenly spaced and consists of Gaussian noise, the Lomb-Scargle periodogram reduces to the classical periodogram. The Lomb-Scargle periodgram is more computationally efficient than the classic periodogram. Another distinct benefit of Lomb-Scargle peridogram is that the unnormalized periodogram in Equation \ref{eq:9} follows a $\chi^2$ distribution with two degrees of freedom when the error terms are Gaussian noise \citep{scargle}. }


\subsection{Spline and Smoothing Methods}
\subsubsection{Cubic Spline}
General cubic spline methods are  described in \citet{wasserman}. Here we use the definitions and notation from \citet{reimann1994}, which are specifically tailored for periodic data. 

A function $s(\cdot)$ on interval $[0,1]$ is a periodic cubic spline with $K$ knots at $t_k$, where $k=1,2,\cdots, K$. It should satisfy the following properties:
\begin{itemize}
\item In each interval $[t_{k-1}, t_k]$, $k=1,2,\cdots, K+1$ (Define $t_0=0$ and $t_{K+1}=1$), $s(\cdot)$ is a polynomial of degree three.
\item $s(\cdot)$ and its first and second order derivatives are continuous everywhere in $[0,1]$ and satisfy the periodicity constrains
\begin{equation*}
    s^{(l)}(0) = s^{(l)}(1), \qquad l=0,1,2.
\end{equation*}
\end{itemize}
Given the set of knots, the spline model is determined by coefficients in each interval satisfying the above constraints. The model is fitted to data using least squares.  To find the period using a cubic spline:
\begin{itemize}
\item convert the raw data $(t_n, Y_n)$ into the phased data $(\rho_n, Y_n)$ by $\rho_n =  \frac{t_n}{p} \, mod \, 1$ for a trial period $p$ (thus the phase space $\rho \in [0,1]$), where $n=1,2,\cdots, N$; 

\item fit a cubic spline for the phased data $(\rho_n, Y_n)$ for the fixed number of knots $K$ for all trial periods, and compute the corresponding $\text{SSE}(p)$. 
\end{itemize}
The estimate $\hat{p}$ is the period that minimizes $\text{SSE}(p)$. 
    

\subsubsection{Local Linear Regression and Supersoomther}
Local linear regression uses weighted averaging to provide a linear approximation to a  nonlinear function at a point \citep{wasserman}. To estimate the period, fit a local linear regression on the phased data $(\rho_n, Y_n), \, n=1,2,\cdots, N$ for some bandwidth. Let $B_i$ denote the $i^{th}$ bandwidth and $J$ the number of observations in $B_i$. We use fit a local linear regression in each band

\begin{equation*}
Y_j = \alpha + \beta \rho_j + \epsilon_j, \qquad \rho_j \in B_i,\; j=1,2,\cdots, J,
\end{equation*}
where $\epsilon_j$'s are i.i.d. error terms. The local linear estimator in each band can be computed by (weighted) least squares
\begin{equation*}
\hat{Y}_j = \hat{\alpha} + \hat{\beta} \rho_j, \qquad \rho_j \in B_i,\; j=1,2,\cdots,J,
\end{equation*}
where $\hat{\alpha}$ and $\hat{\beta}$ are obtained from local fits to data points in each band. The estimate $\hat{p}$ is the period that minimizes $\text{SSE}(p)$. 


Friedman's Supersmoother \citep{frie} performs three linear smooths of the phased data $(\rho_n, Y_n)$, $n=1,2,\cdots, N$ with long, medimum and short bandwidths. Then it does a local cross-validation to determine which bandwidth gives the best fit at each phase value. The period estimate is obtained through minimizing the \textbf{Sum of Absolute Residuals} (SAR)
\begin{equation}
\text{SAR}(p) = \sum_{n=1}^N \frac{1}{\hat{\sigma}_n} |Y_n - \hat{Y}_{n}(p) |,
\end{equation}
where the $\hat{Y}_{n}(p)$ is the fitted value from Supersmoother at a trial period $p$, and $\hat{\sigma}_n$ is the estimate of the standard deviation of the errors.

\subsection{Phase-folding Methods}
\subsubsection{String Length Methods}
Phase-folding methods compute the dispersion of the data in the phase space to search the period that minimizes the dispersion. For example, String-length computes the phase of the raw observations for each trial period $p$ and sorts phase data in an ascending order of the phase. Let $(\rho_n^*, Y_n^*)$ be the ordered phased data, where $n=1,2,\cdots, N$. The best period minimizes the String-length statistic
\begin{equation}
\text{SL}(p) = \sum_{n=1}^N \left[  (Y_{n+1}^* - Y_n^*)^2 +  (\rho_{n+1}^* - \rho_n^*)^2 \right].
\end{equation}
Note $Y_{N+1}^* = Y_{1}^*$ and $\rho_{N+1}^* = \rho_{1}^*$.

However, String-length depends on the differences in the phase as well as in the response, so a change in either could lead to a different estimate of the period. \citet{lk} recommended minimizing the following statistic
\begin{equation}
\text{LK}(p) = \sum_{n=1}^N (Y_{n+1}^* - Y_n^*)^2.
\end{equation}

Another modified string length method is due to \citet{renson} which estimates the period by minimizing the quantity:
\begin{equation}
\text{REN}(p)=\sum_{n=1}^N \frac{(Y_{n+1}^* - Y_n^*)^2}{(\rho_{n+1}^* - \rho_n^*)^2+b^2},
\end{equation}
where $b$ is chosen so that the difference $(\rho_{n+1}^* - \rho_n^*)^2+b^2$ won't be too small.

\subsubsection{Phase Dispersion Minimization}
The variance is computed by
\begin{equation*}
\sigma^2 =\frac{1}{N} \sum_{n=1}^N (Y_n -\bar{Y})^2, \quad \text{where $\bar{Y} = \frac{1}{N} \sum_{n=1}^N Y_n$.}
\end{equation*}
If we divide the data into $M$ distinct samples and each sample has $n_m$ observations, where $m=1,2,\cdots, M$, then each sample has the variance $s_m^2$, and the overall variance of all samples is 
\begin{equation}
s^2 = \frac{\sum_{m=1}^M(n_m-1)s_m^2}{\sum_{m=1}^M n_m -M}
\end{equation}

The Phase Dispersion Minimization method(PDM) is implemented in two steps: 

\begin{itemize}
\item convert the raw data into the phased data for a trial period $p$ by $\rho_i = \frac{t_i}{p} \, mod \, 1 $, where $i=1,2,\cdots, N$. $\rho \in [0,1]$;
\item divide the full phase interval $[0,1]$ into $M$ fixed bins with observations in each bin chosen so that these observations have similar phase. 
\end{itemize}
Compute the PDM statistics for each trial period $p$ by

\begin{equation}
\text{PDM}(p) = \frac{s^2(p)}{\sigma^2} =\displaystyle \frac{\frac{\sum_{m=1}^M(n_m-1)s_m^2}{\sum_{m=1}^M n_m -M}}{\frac{1}{N} \sum_{n=1}^N (Y_n -\bar{Y})^2}.
\end{equation}


If $p$ is not the correct period, then $s^2(p) \approx \sigma^2$ and PDM$(p) \approx 1$; if $p$ is the correct period, then PDM$(p) $ will reach a local minimum compared with the neighboring periods, ideally near zero.

\subsection{Statistical Inference on the Period}
We briefly discuss confidence intervals for the period.  The finite sample distribution of the estimated period is not generally quantifiable.  However, under restrictive assumptions some exact distributional results are available.  For example, as noted above under the assumption of normality the unnormalized Lomb-Scargle peridogram  in Equation (\ref{eq:9}) follows a $\chi^2$ distribution with two degrees of freedom.  This can be used to make a confidence interval \citep{Baluev}.  More generally applicable procedures could be based on asymptotic (large sample) normality.  Since the Lomb-Scargle peridogram solves a least squares minimization, asymptotic normality can be established and used to create an approximate interval.  Similarly, the smoothing methods (spline and local), result in estimators which have asymptotic normal properties \citep{wasserman}.  In practice, we recommend using bootstrapping to estimate standard errors and create confidence intervals.  The interested reader is again referred to \cite{wasserman} for details.  In the current paper, we seek to provide point estimates of the period, so we do not pursue this further here.  Rather we provide the practitioner some guidance to select the most reliable method by conducting a simulation study in the next section.  

\section{A Brief Comparison of Period Finding Methods}\label{compare}
\citet{graham} has conducted a comparison of several period finding techniques applied to observational data of variable stars from three projects: Catalina Real-time Transient Survey, ASAS Catalog of Variable Stars and MACHO. {However, his comparison did not consider the heterogeneity in data.  Moreover, the true period of a variable star is actually decided by computation instead of prior knowledge, so the comparison of accuracy measures is not persuasive. Therefore, we present a different comparison by simulating the sea level data at the La Jolla Station, California.} The seal level data were collected by the project ``Permanent Service for Mean Sea Level''. The periodic variation of the sea level is known to be annual because of the ``steric effect'', which is caused by the annual variation in water temperature at shallow depths. {The sea level data consist of $300$ observations, which ranges from 1992 to 2017 and are evenly sampled. The sampling rate is $12$, i.e., $12$ observations per year, the time scale has been modified to be month/12 so that the natural period is 1, and the standard error of the white noise in the data is estimated to be $\sigma = 90.53$ millimeter.} 

{Our goal is to examine how period estimation methods behave under specified non-uniformity of measurement times and differing variability of additive random noise. To simulate the non-uniformity, we mimic missing data by randomly selecting time points and remove the observations which are not selected.  We vary the proportion of sampling observations from $20\%$ to $70\%$.   With a lower proportion of the data being randomly sampled, the sampled data become more unevenly spaced. For each selected proportion $prop$, we randomly select $300 \times prop$ time points to sample, then estimate the period using each of the methods described in this paper.  We replicate this $100$ times, thus creating $100$ samples for each proportion.  We quantify the performance of the different methods in two ways.  First, we consider the classical statistical mean squared error (MSE):
\begin{equation}
\frac{1}{100} \sum_{i=1}^{100} (\hat{p}_i-p_0)^2,
\end{equation}
where $\hat{p}_i$ is the estimated period from artificial data $i$, and $p_0$ is the known period.}  As a second measure of performance we consider the \textbf{accuracy metric} from \citet{oluseyi} which has been also used by \citet{graham} in his comparison:
\begin{align}
    \frac{|\hat{p}-p_0|}{p_0} &\le \frac{\delta \phi_{\max}p_0}{\Delta \tau},
\end{align}
where $p_0$ is the true period, $\Delta \tau$ is the duration of the time series, $\delta \phi_{\max}$ is the maximum allowed phase offset after period-folding some cycles. In sea level data $p_0=1$ year, $\Delta \tau=25$ years, also considering that the minimum spacing between two trial periods is $0.005$ year in the simulation, we simplify the above accuracy metric to 
\begin{align}
    |\hat{p}-1| \le 0.01 \text{ year}.
\end{align}
So if the estimate $\hat{p}$ is within  $0.01$ year to the true period $p_0$, we accept $\hat{p}$ as an accurate estimate. The percentages of accurate estimates of each method in $100$ runs are plotted as follows:
\begin{figure}[H]
\centering
\includegraphics[trim={0 0 0 4cm}, scale=0.45]{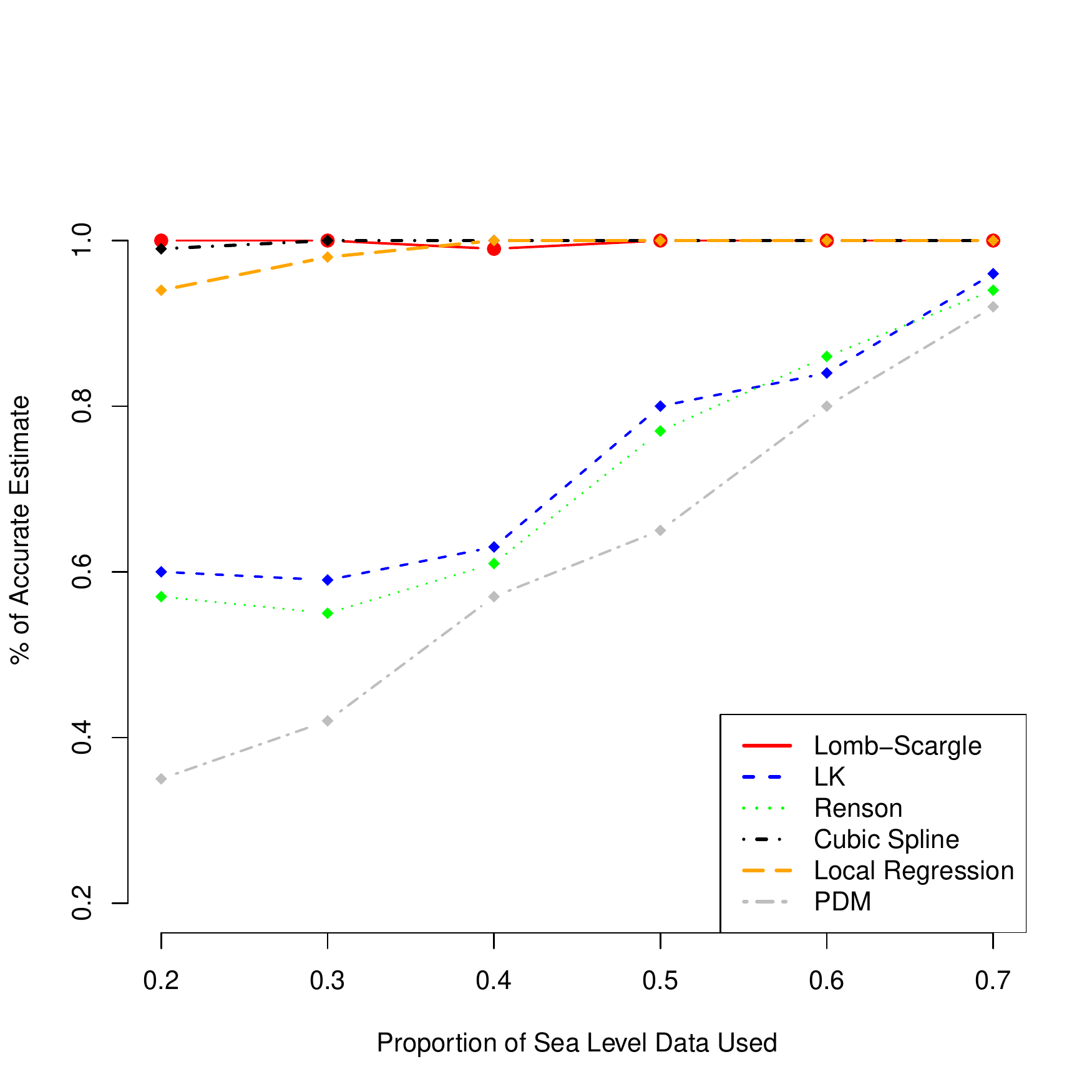}
\caption{Accuracy comparison of period finding methods. Left side of graph corresponds to more uneven spacing.}
\label{fig:propaccu}
\end{figure}
The accuracy in Figure \ref{fig:propaccu}  and mean square error in Table \ref{tab:mse_prop} suggest that Lomb-Scargle periodogram and cubic spline have a robust performance in every unevenly spaced case (different proportions sampled). However, Lomb-Scargle periodogram has slightly smaller mean squared error than the cubic spline. Local regression has the smallest MSE when 40\% or more of sea level data are sampled. The  Lafler-Kinman(LK), Renson(REN)  and Dispersion Minimization(PDM) methods perform poorly relative to the other three methods. 


\begin{table}[H]
\footnotesize
    \centering
        \begin{tabular}{c c c c c c c c}
            \toprule
            \midrule
                && \multicolumn{6}{c}{MSE When Different Proportion of Data Randomly Sampled}\\ \cmidrule{3-8}
                && $20\%$  & $30\%$ & $40\%$ & $50\%$ & $60\%$ & $70\%$  \\ \cmidrule{3-8}
                \multicolumn{1}{c}{\multirow{6}{*}{\begin{sideways}\parbox{2cm}{\centering Methods}\end{sideways}}}   &
                \multicolumn{1}{l}{Lomb-Scargle}& $1.66\times 10^{-5}$ & $1.42\times 10^{-5}$ & $7.16\times 10^{-6}$ & $7.04\times 10^{-6}$ & $3.73\times 10^{-6}$ & $3.82\times 10^{-6}$   \\
                \multicolumn{1}{c}{}    &
                \multicolumn{1}{l}{Cubic Spline} & $1.74\times 10^{-5}$ & $1.19\times 10^{-5}$ & $9.69\times 10^{-6}$ & $8.44\times 10^{-6}$ & $7.78\times 10^{-6}$ & $6.9\times 10^{-6}$  \\
                \multicolumn{1}{c}{}    &
                \multicolumn{1}{l}{Local Regression} & $0.12$ & $0.02$ & $6.85\times 10^{-6}$ & $4.05\times 10^{-6}$ & $3.64\times 10^{-6}$ & $3.07\times 10^{-6}$  \\
                &\multicolumn{1}{l}{LK}& $1.16$ & $1.06$ & $0.98$ & $0.62$ & $0.38$ & $0.05$  \\
                \multicolumn{1}{c}{}    &
                \multicolumn{1}{l}{Renson} & $1.32$ & $1.12$ & $1.03$ & $0.71$ & $0.36$ & $0.13$  \\
                \multicolumn{1}{c}{}    &   
                \multicolumn{1}{l}{PDM} & $1.88$ & $1.50$ & $1.15$ & $0.95$ & $0.44$ & $0.20$   \\
            \midrule
            \bottomrule
        \end{tabular}
        \centering
        \caption{Random selection of  sea level data.}
        \label{tab:mse_prop}
\end{table}

{As data gets noisier, period finding methods perform worse and eventually fail to detect the true period. Thus, the resistance of each method to increased variability of white noise is of great interest. To examine the impact of noisy data, we randomly sample $60\%$ of the sea level data $100$ times, in each case we add additional Gaussian random noise with different standard deviation at $0.5\sigma, 1.0\sigma, 1.5\sigma, 2.0\sigma, 2.5\sigma$ to the sampled data.} Note $\sigma$ is the estimated standard deviation of the original sea level data, so adding noise at $0.5\sigma$ increases the total noise variance by 1.25 relative to the original data variance, while adding noise with standard deviation $2\sigma$ increase total variance by a factor of 5. 

\begin{figure}[H]
\centering
\includegraphics[trim={0 0 0 3.5cm}, scale=0.5]{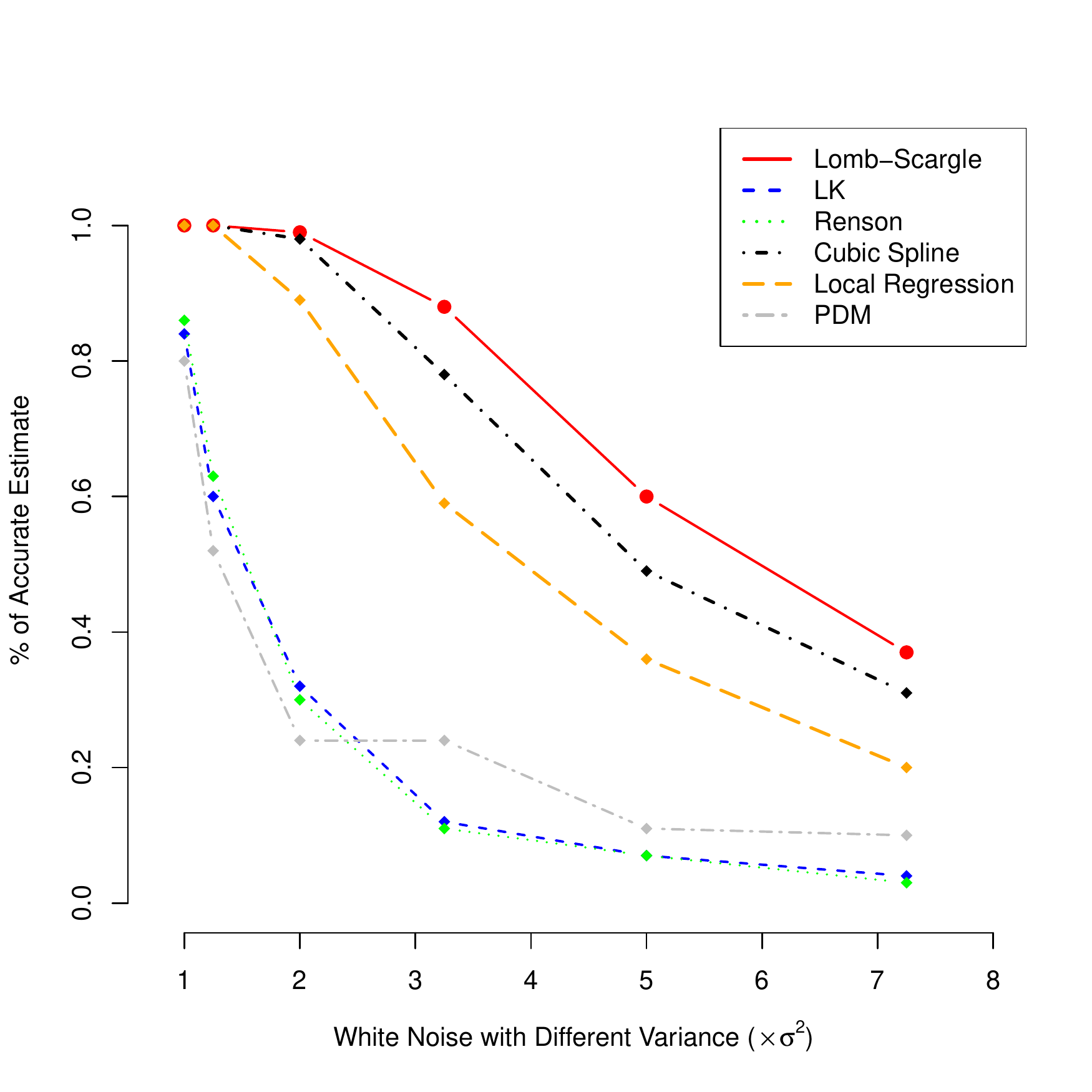}
\caption{Accuracy of period finding methods under increased noise.  The values on $x-$axis are the multiples of variance of original sea level data. }
\label{fig:wnaccu}
\end{figure}

The accuracy plot in Figure \ref{fig:wnaccu} and MSE in Table 2 suggests that the performance of Lomb-Scargle periodogram and cubic spline deteriorate sharply when the variance of the white noise is above $2\sigma^2$. The misleading estimates of Lomb-Scargle periodogram in noisy unevenly spaced time series in the simulation coincide with the partial findings from \citet{schimmel}.

To summarize, our simulations seem to indicate that Lomb-Scargle periodogram and cubic spline are most reliable in finding the period. 

\begin{table}[H]
    \centering
        \begin{tabular}{c c c c c c c c}
            \toprule
            \midrule
                && \multicolumn{6}{c}{MSE Under Different White Noise}\\ \cmidrule{3-8}
                && $\sigma^2$  & $1.25\sigma^2$ & $2\sigma^2$ & $3.25\sigma^2$ & $5\sigma^2$ & $7.25\sigma^2$\\ \cmidrule{3-8}
                \multicolumn{1}{c}{\multirow{6}{*}{\begin{sideways}\parbox{2cm}{\centering Methods}\end{sideways}}}   &
                \multicolumn{1}{l}{Lomb-Scargle}& $3.73\times 10^{-6}$ & $6.45\times 10^{-6}$ & $1.01\times 10^{-5}$ & $0.96$ & $22.6$ & $14.7$ \\
                \multicolumn{1}{c}{}    &
                \multicolumn{1}{l}{LK}& $0.39$ & $0.73$ & $0.79$ & $0.82$ & $0.78$ & $0.65$ \\
                \multicolumn{1}{c}{}    &
                \multicolumn{1}{l}{Renson} & $0.36$ & $0.73$ & $0.83$ & $0.94$ & $0.83$  & $0.65$\\
                \multicolumn{1}{c}{}    &   
                \multicolumn{1}{l}{Cubic Spline} & $7.78\times 10^{-6}$ & $1.00\times 10^{-5}$ & $2.04\times 10^{-5}$ & $0.26$ & $0.64$ & $0.68$ \\
                \multicolumn{1}{c}{}    &
                \multicolumn{1}{l}{Local Regression} & $3.64\times 10^{-6}$ & $7.37\times 10^{-6}$ & $0.25$ & $0.76$ & $0.91$  & $0.94$ \\
                \multicolumn{1}{c}{} &
                \multicolumn{1}{l}{PDM} & $0.44$ & $1.29$ & $2.23$ & $1.68$ & $1.37$ & $1.14$ \\
            \midrule
            \bottomrule
        \end{tabular}
        \centering
        \caption{Mean squared error of period finding methods under white noise with different variance.}
        \label{tab:mse_wn}
\end{table}


\section{An Application:  Periodicity in the Light Magnitude of Variable Stars}\label{application}

Determination of the periodicity is a fundamental issue in the study of variable stars, which includes classification of variable stars, calibration of the period-luminosity relation, determination of the pulsation modes, detection of stellar rotation and so on.

The data of the variable stars were collected through MACHO project, which is a collaboration of scientists at the Mt. Stromlo and Siding Spring Observatories, the Center for Particle Astrophysics at Santa Barbara, San Diego, the University of California at Berkeley, and the Lawrence Livermore National Laboratory. Data were collected daily over a 4-year period when weather permitted, on approximately 8 million stars in the Large Magellanic Cloud (LMC) and the bulge of the Milky Way. 

A Cepheid variable is a type of star that pulsates varying in both diameter and temperature and producing changes in brightness with a well-defined stable period and amplitude. Data of Lcb1 Cepheid variable star from LMC is used as the example which consists of $327$ observations made in $385$ days. The maximum spacing(gap) between two observations is $32.83$ days. 


\begin{table}[H]
 \centering
 \begin{tabular}{l c}
 \hline
 Methods &  Period Estimate (in days) \\ 
 \hline
    Classical Periodogram & 12.86* \\
    Lomb-Scargle Periodogram & 13.14\\    
    Lafler-Kinman & 13.12 \\ 
    Renson & 13.12  \\ 
    Cubic Spline & 13.16  \\ 
    Local Regression &  13.12 \\ 
    Phase Dispersion Minimization & 13.13 \\
 \hline
 \multicolumn{2}{l}{}
 \end{tabular}
 \caption{Period Estimate of Lcb1 Variable Star. }
 \label{tab:lcb1}
 \end{table}

\begin{figure}[H]
\centering
\includegraphics[trim={0 0 0 1.5cm}, scale=1]{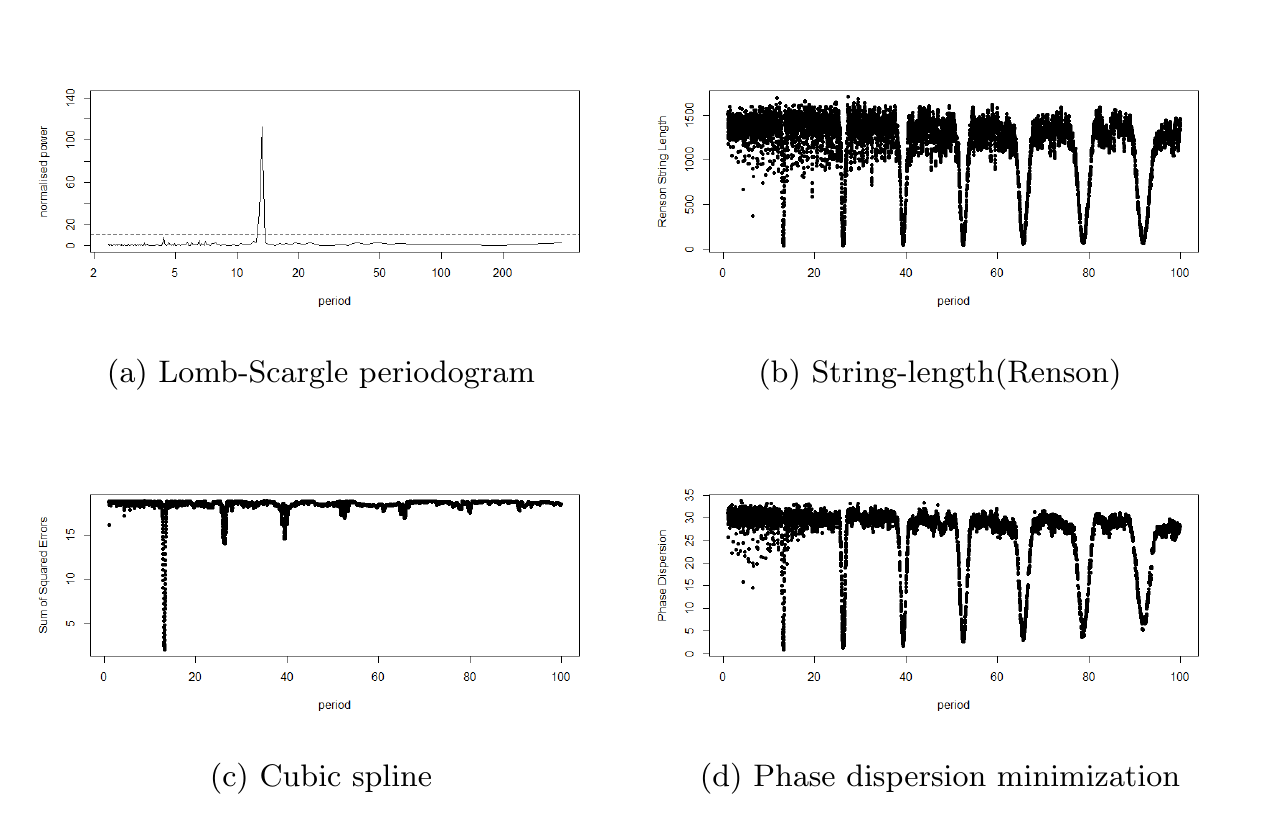}
\caption{Period Estimate of Lcb1 Variable Star using Different Period Finding Methods. }
\label{fig:lcb1}
\end{figure}

{
Table \ref{tab:lcb1} lists the period estimates of Lcb1 variable star by different algorithms, while Figure \ref{fig:lcb1} visualizes how each method searches for the period. Note that Lafler-Kinman and Local Regression were not plotted since their search paths are identical to Renson and cubic spline.} 

{The horizontal dash line in plot(a) Lomb-Scargle periodogram is the critical value. Periodogram peaks exceeding this line are considered significant and thus is the estimate of period. The significant level $\alpha=0.01$ is used. Renson method in plot(b) calculates the string-length statistic for all trial periods, the minimum statistic, usually the first valley, is taken as the period estimate. Cubic spline and Phase disperson minimization are similar to Renson string-length but minimize different statistics. There are multiple valleys in Renson string-length and Phase dispersion minimization plots other than the first. They correspond to the multiple of the period and may have close statistics to the first valley. String-length methods and Phase Dispersion minimization are less likely to distinguish these valleys and thus estimate multiple periods when data are noisier or more irregularly spaced. }




%
%
%
%

\section{Conclusions}
Cyclic behavior is a prominent feature in many types of geophysical data.  These periodic effects can be estimated without specifying a parametric model, and should be accounted for in statistical analysis of the data.   In this paper we consider estimating unknown cycle lengths in non-uniform time series data. We have examined their performance by a simulation on sea level data. Then  we have considered the case of estimating a primary period as in the case of the light magnitude of a periodic star. Several of the methods in this paper work quite well in this case, especially Lomb-Scargle periodpgram and periodic cubic spline.


As a final comment we note that many geophysical time series may have multiple cycles impacting the data. Modern statistical model selection methods such as LASSO can be used to simultaneously determine the important periods and estimate the effects at each important period \citep{lasso}. Other unsolved research problems still require efforts, for example, unevenly spaced data that consist of periodic signals with non-sinusoidal shapes, or correlated noise.

\section{Appendix: Hardware and Software Specifications}
Hardware:
\begin{itemize}
\item[1.] CPU: Intel Core i7-8700K 3.70 GHz
\item[2.] Memory: 16GB DDR4-2400 MHz 
\item[3.] GPU: Nivdia GTX 1060 
\end{itemize}

Software:
\begin{itemize}
\item[1.] Operating System: Windows 10 Professional
\item[2.] R Version 3.3.2 
\end{itemize}

\begin{acronyms}
\acro{i.i.d.}
independent identically distributed
\acro{SSE}
Sum of Squared Errors
\acro{MSE}
Mean Squared Errors

\acro{LK}
Lafler-Kinman's String Length

\acro{REN}
Renson's String Length

\acro{SAR}
Sum of Absolute Residuals

\acro{PDM}
Phase Dispersion Minimization
\acro{LASSO}
Least Absolute Shrinkage and Selection Operator
\end{acronyms}

\acknowledgments
{We thank the referees and the associate editor for their meticulous reading of our original manuscript and their useful comments. The paper has made use of data obtained by MACHO project and ``Permanent Service for Mean Sea Level'' project.
}

\bibliography{agusample}

\end{document}